# An all-silica three-element wide-field corrector for GMT

Will Saunders[1*], Peter Gillingham[1], Sean Lin[2], Bob Woodruff[2], Andrew Rakich[2]

[1] Australian Astronomical Observatory, PO Box 915, North Ryde, NSW 1670, Australia

[2]Giant Magellan Telescope Organisation, 465 N Halstead St #250, Pasadena, CA 91107, USA

## ABSTRACT

We present an alternative Corrector-ADC design for GMT. The design consists of just 3 silica lenses, of maximum size 1.51m, and includes only a single low-precision asphere for 20' field-of-view, and none for 10'. The polychromatic (360nm-1300nm) image quality is $d_{80}$<0.043" at zenith and $d_{80}$<0.20" for ZD<60 degrees. The monochromatic image quality is $d_{80}$<0.1" everywhere, and typically ~0.05″. The ADC action is achieved by tilt and translation of all three lenses; L1 and L2 via simple slide mechanisms each using a single encoded actuator, and L3 via a novel 'tracker-ball' support and three actuators. There is also a small motion of M2 via the hexapod, automatically generated by the AGWS system. The ADC action causes a small non-telecentricity, but this is much less than the unavoidable chromatic effects shared with the baseline design. The ADC action also changes the distortion pattern of the telescope, but this can be used positively, to reduce the maximum image motion due to differential refraction by a factor of three. The transmission is superb at all wavelengths, because of the reduced number of air/glass surfaces, and the use only of fused silica.

**Keywords:** Telescope optical designs, wide-field corrector, atmospheric dispersion corrector, atmospheric dispersion correction, wide-field spectroscopy, multi-object spectroscopy, Giant Magellan Telescope, Extremely Large Telescopes.

## 1. INTRODUCTION

The Giant Magellan Telescope (GMT) [1] is designed to have a wide-field capability, with a Corrector-ADC (C-ADC) giving 20′ field-of-view, to feed the MANIFEST fiber-feed front-end [2], or a 10′ FoV to feed an imaging spectrograph [3]. The baseline C-ADC design has a total of 6 elements, including an Atmospheric Dispersion Corrector (ADC) consisting of a pair of counter-rotating Risley prismatic crown/flint doublets. These are of ~1.5m size, and homogeneity cannot be guaranteed at this scale, so post-polishing is required to correct for glass inhomogeneity.

A radically different C-ADC design is offered by the Compensating Lateral ADC (CLADC) concept, which allows an ADC action for almost any Wide-Field Corrector (WFC), with no additional lenses [4,5,6]. The idea is that translating a lens laterally introduces a variable prismatic effect, which can correct for atmospheric dispersion, but adds wavefront tilt and astigmatism; translating one or more other lenses in the system can correct for the tilt and astigmatism, leaving a pure ADC action. The concept works for both prime focus and two-mirror telescopes, and (at least in principle) has no impact on either throughput or monochromatic image quality. This is especially useful for any design working over a wide wavelength range, because both Risley/Amici or Subaru-style [7] ADCs give a large throughput penalty in the UV and NIR, because of glass absorption, and because of the additional air/glass surfaces used at the limits of their coating design ranges. The wavelength range used in the optimization of this design was 360nm-1300nm, but it works acceptably from the limit of the UV atmospheric cutoff to the thermal IR.

For the GMT, the placement of C-ADC lenses is constrained by design decisions already taken for the telescope geometry. The CLADC concept requires that (at least) two lenses be moved, with the largest possible separation between them. In practice, this means that the final C-ADC element, close to the focal plane, must be moved. This is awkward because it is below the instrument derotator, while the required motion is meridional. However, a simple and robust mechanism for the required motion has been developed as part of the 3dF project for the AAT [8], since equatorial telescopes have the same problem.

* will@aao.gov.au

The design also demands a significant movement of M2, to correct for higher-order aberrations introduced by the lens motions. These motions are well within the physical movement limits of the hexapod, especially as they are in the same sense as the flexure of the telescope top-end with gravity.

This moving-lens style of ADC action has some features not present in Amici-style ADCs.

1. If the lens diameters are restricted to that required at Zenith, the lens motions cause vignetting for some parts of the field at non-zero ZDs. Subaru-style ADCs also have this feature, even more strongly.

2. The motion of the lenses introduces changes to the distortion pattern at the focal surface, of similar magnitude to those caused by differential refraction, but with a more complex pattern. However, the ADC action also readily allows small changes of plate-scale. The combined effect can be tuned to give a large reduction in the overall image motion at the focal plane, compared with uncorrected differential refraction.

3. The lens motion increases the non-telecentricity at the focal surface, but for the design presented here, the effect is very small compared with irreducible chromatic effects caused by L3 being a strongly powered singlet.

4. The ZD-dependent non-telecentricity also requires significant changes to the operation, though not the design, of the Acquisition, Guide, and Wavefront Subsystem (AGWS).

## 2. DESIGN DETAILS

At zenith setting, the design presented here looks somewhat like the baseline design, with a positive L1, a negative L2, and a strongly positive field lens L3 (Figure 1), and with a FoV of 20′. The distance from the vertex of M1 to L1 has been increased, to reduce the diameter of L1. The basic design is all-silica, with L1 and L2 spherical. Their diameters, including 30mm for mounting, are 1510mm and 1440mm. For 20′ FoV, L3 has diameter 1340mm and is aspheric on its rear surface, with maximum aspheric departure 21mrad. There is a large tolerance in the slope error, because the surface is so close to focus. The focal surface has diameter 1281mm and ROC 3276mm. The WFNO is f/8.28 (vs baseline f/8.34), focal length is 210.9m, plate scale on-axis 1.027mm/″, maximum distortion 1.4%.

For 10′ FoV, L3 is spherical with diameter 688mm and thickness 80mm. The focal surface has diameter 629 mm with ROC 3177mm. The WFNO is f/8.28, focal length is 211.7m, plate scale on-axis 1.022mm/″, maximum distortion 0.35%.

The ADC action is implemented by moving laterally and tilting all three lenses, as shown in Figure 1. At ZD=60°, the lateral motions of L1 and L2 are 62mm, with tilts of 2.8°. For L3, the maximum motion is 37mm and 1°. At other ZDs, the required motions are varied according to the standard formula for atmospheric refraction[1].

As discussed further below, the active optics system (AcO) will respond to the global aberrations the ADC action introduces (focus, astigmatism, coma), by demanding movements of M2. For ZD 60°, these consist of (a) a 7.2mm translation (downwards, so not increasing the total required range of travel), (b) a tilt of < 11′, and (c) a refocus of up to 20µm.

The design was optimized for 360µm, 370µm, 400µm, 450µm, 550µm, 750µm, 1000µm, 1300µm, with half-weighting for the first and last wavelengths, which are outside the core 370nm-1000nm range. Field positions were strongly weighted towards the periphery, because the image quality changes faster there. Both ±Y fields were used, because the ADC action introduces an asymmetry.

[1] tan((ZD - 10.3°/(95.11° - ZD))

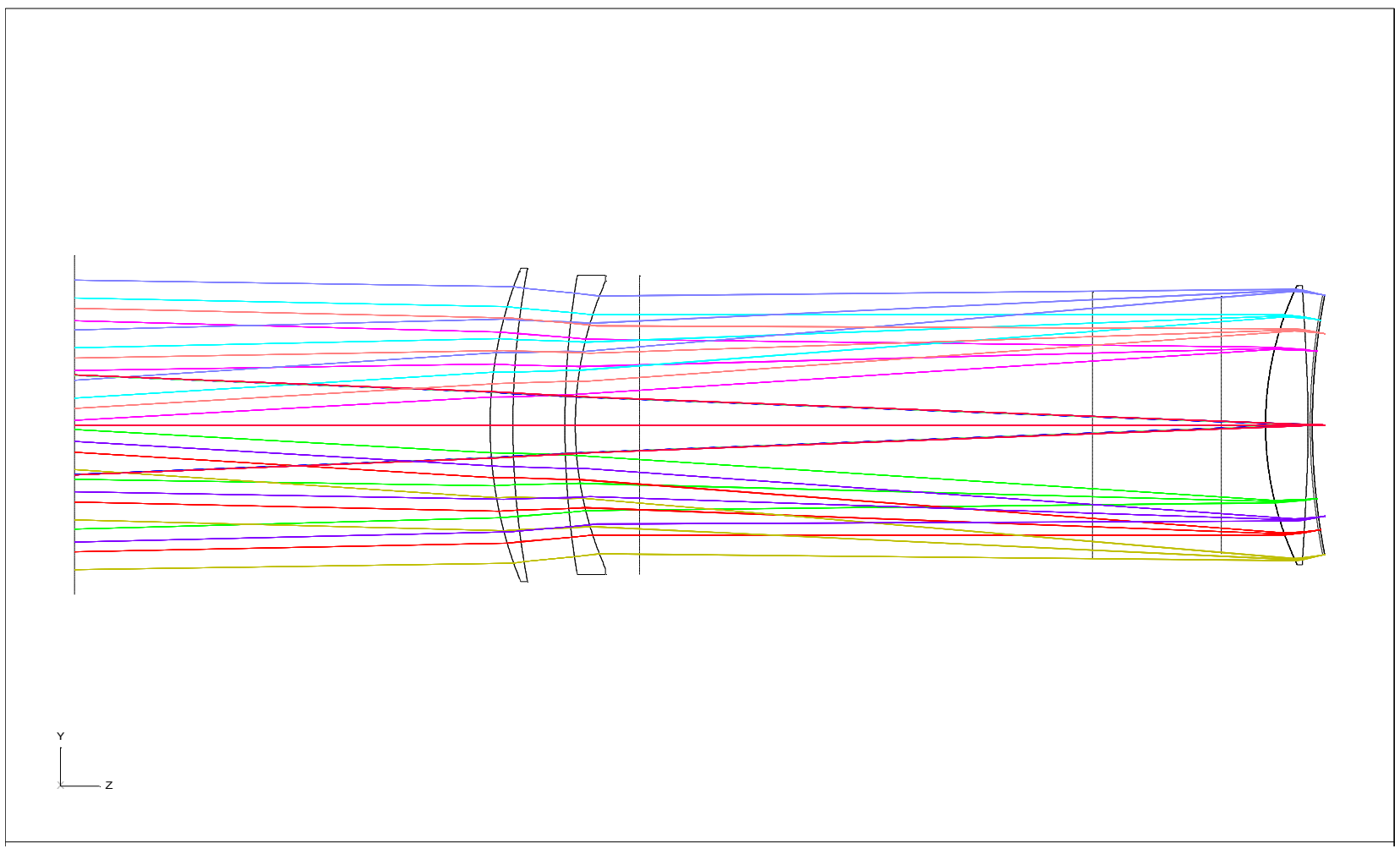

**Figure 1. ZEMAX layout at Zenith. All lenses are fused silica. The 2nd surface of L3 is aspheric with up to 26μm/mm aspheric departure. GMT reference surfaces are also shown (from left to right: bottom of mirror cell, top and bottom of Instrument Platform, top of Gregorian Instrument Rotator instrument volume).**

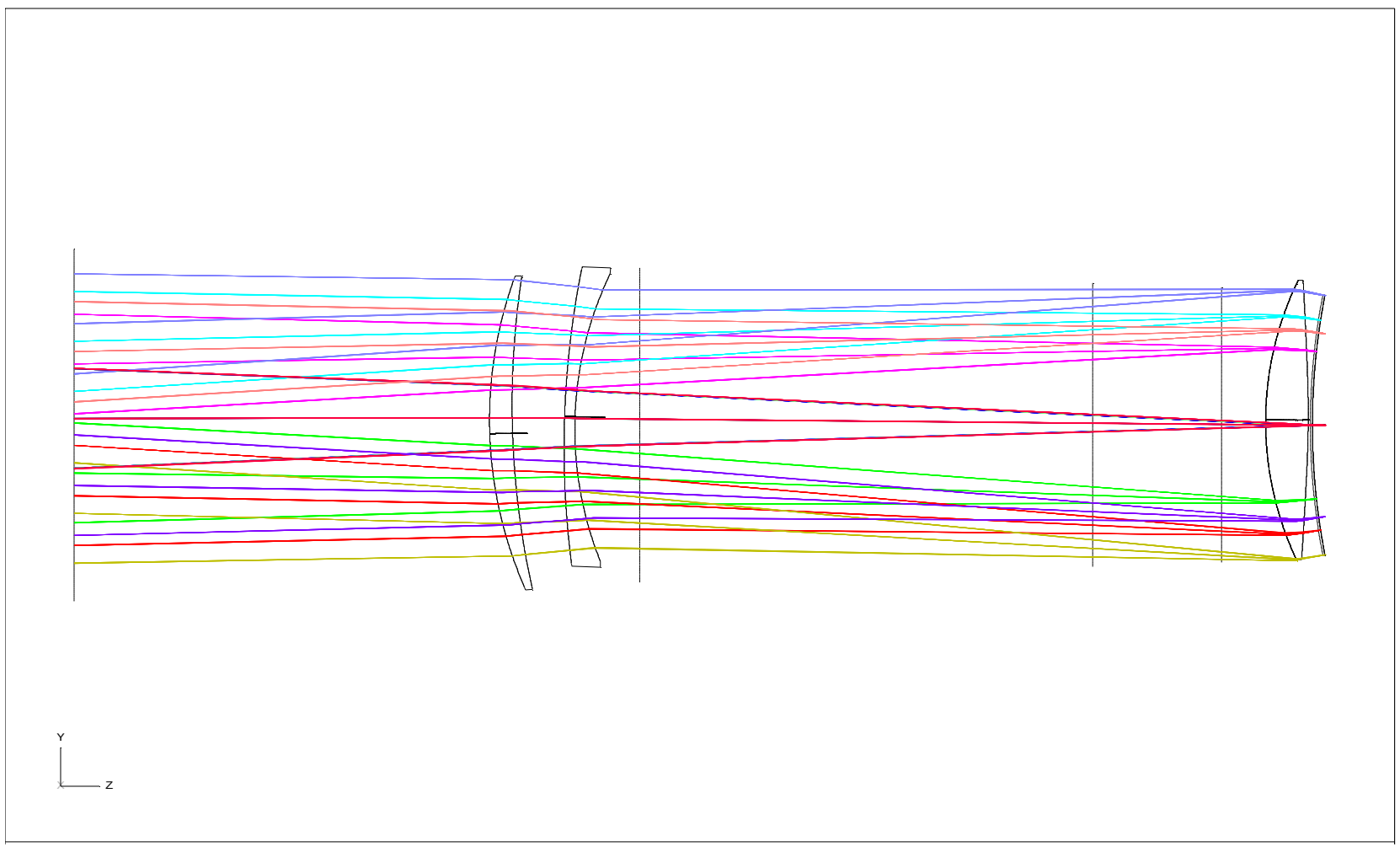

**Figure 2 (b) Layout for ZD=60°, showing displacements of L1 and L2 of ±62mm, with tilts of ±2.8°, and displacement of L3 by +35mm, with a tilt of +0.8°.**

## 3. IMAGING PERFORMANCE

Figure 3 shows spot diagrams for ZD=0 and 50° for 20′ FoV. The imaging performance at Zenith is excellent, though there is inescapable chromatic variation in the image quality in the central parts of the field, meaning the best UV/NIR/polychromatic image quality is at larger radii (Figure 4(a)). Away from Zenith, the monochromatic image quality remains excellent, but there is significant secondary spectrum, because silica has partial dispersion poorly matched to the atmosphere. The worst 80% enclosed energy diameters are $d_{80}$ = 0.044″ and 0.191″ at ZD=0 and 50°. Figure 5 shows the equivalent spot diagrams for 10′ FoV, and Figure 4(b) shows the rms map. The best image quality is at the field center. The worst 80% enclosed energy diameters are 0.033″ and 0.182″ at ZD=0 and 50°. The deterioration in polychromatic image quality with ZD can usually be evaded: with GMACS by observing at the parallactic angle, and for MANIFEST via the extraction software, (since most modes involve image-slicing). Also, for any observation not requiring the full wavelength range, the ADC setting would be adjusted to minimize the secondary spectrum for that observation.

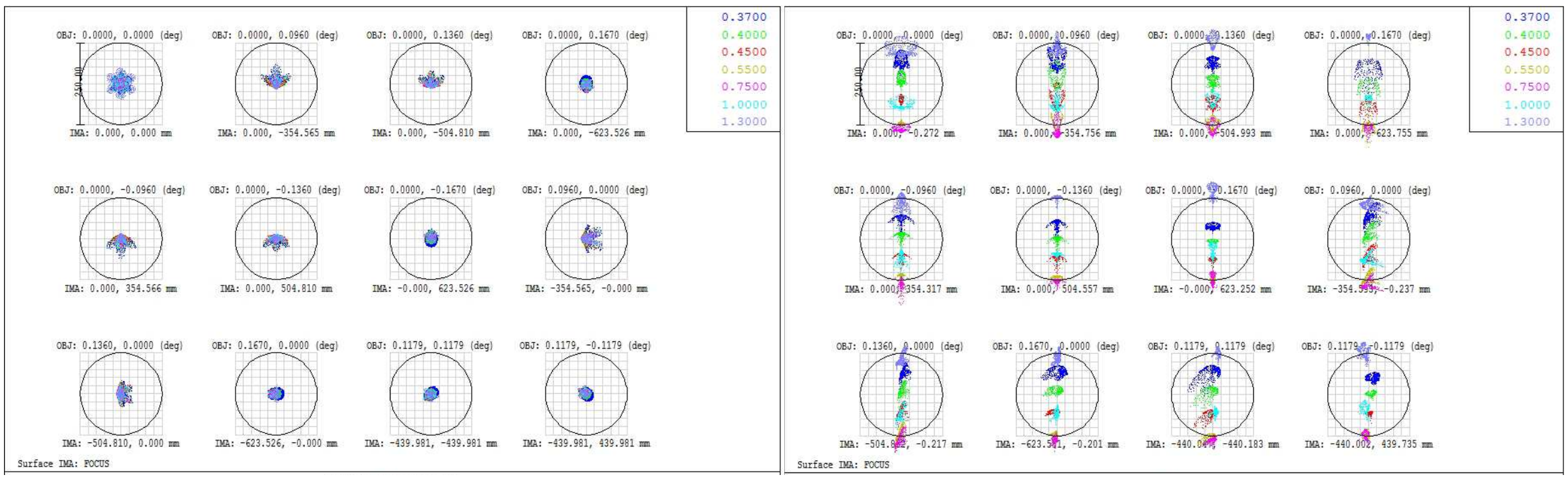

**Figure 3. Spot diagrams for 20′ FoV and 360nm-1300nm, at (a) ZD=0, and (b) ZD=50°. Circle diameter is 250μm (0.245″).**

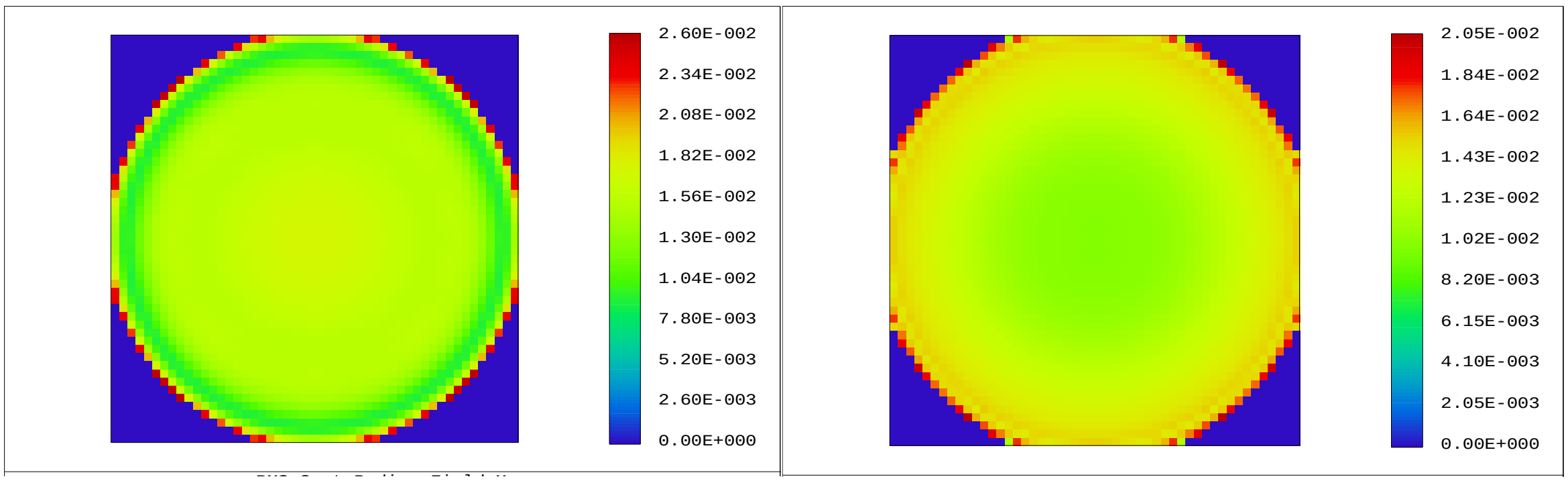

**Figure 4. Polychromatic rms spot radius maps for (a) 20′ FoV and (b) 10′ FoV.**

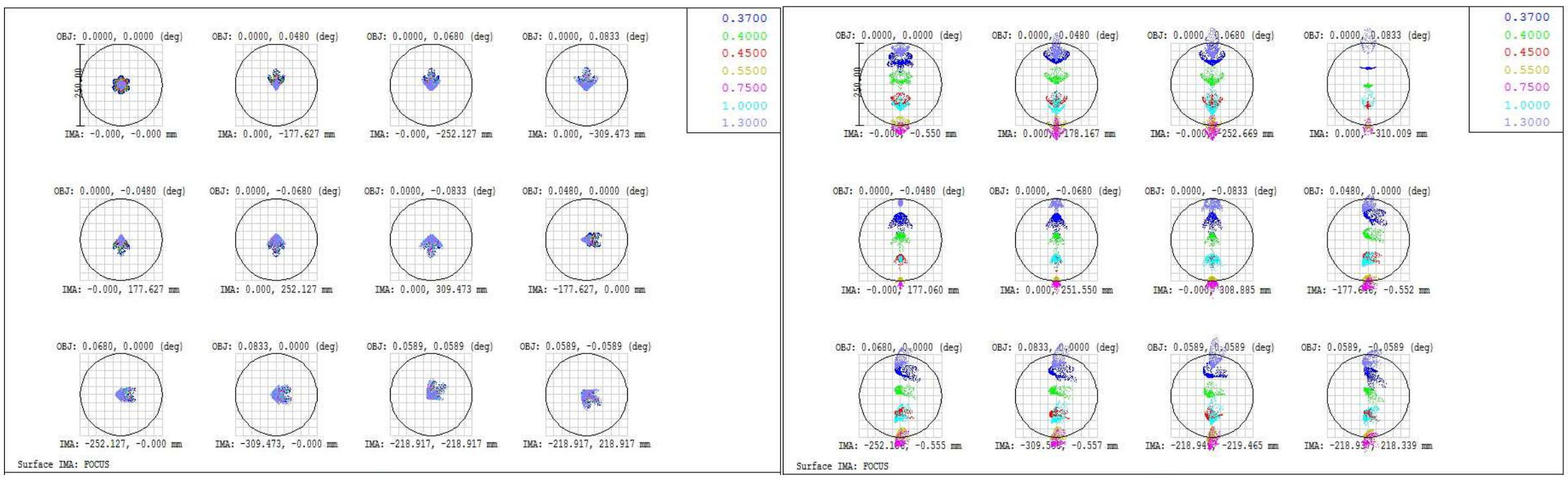

**Figure 5. Spot diagrams for 20′ FoV and 360nm-1300nm, at (a) ZD=0, and (b) ZD=50°. Circle diameter is 250μm (0.245″).**

## 4. VIGNETTING

Total GMT wide-field vignetting losses are dominated by M2 being sized only for on-axis use, causing off-axis mismatch between M1 and M2 apertures. For 20′ FoV, this amounts to 9% at the edge of the field, with an average loss over the whole field of 6%. For a 10′ FoV, the loss is 5% at the edge of the field, with an average over the field of 3%. These losses are independent of the C-ADC design.

For 20′ FoV, our C-ADC design adds ZD-dependent vignetting at L1 and L3 (the beam hardly moves with respect to L2). The L1 vignetting is restricted to a crescent at the lowest part of the field, and is larger in the IR, affecting 10% of the field at ZD=60° (Figure 6). The maximum vignetting anywhere in the field is 3% at ZD=45° and 12% at ZD=60°, with averages over the field of 0.3% and 1%. For L3, the vignetting cuts in very sharply, but only affects 2% of the field at ZD=45° and 3.5% at ZD=60°. These losses can be seen directly as the clipping of the beam for fields #4 and #7 in Figure 9. These vignetting losses could be avoided by increasing the sizes of L1 and L3.

For a 10′ FoV, there is no vignetting at L1 or L2. If L3 is sized for on-axis use, there is vignetting of the upper 4.2% of the field at ZD=45°, and 7.3% at ZD=60° (Figure 10). WFC vignetting could be avoided entirely by increasing the diameter of L3 from 686mm to 726mm.

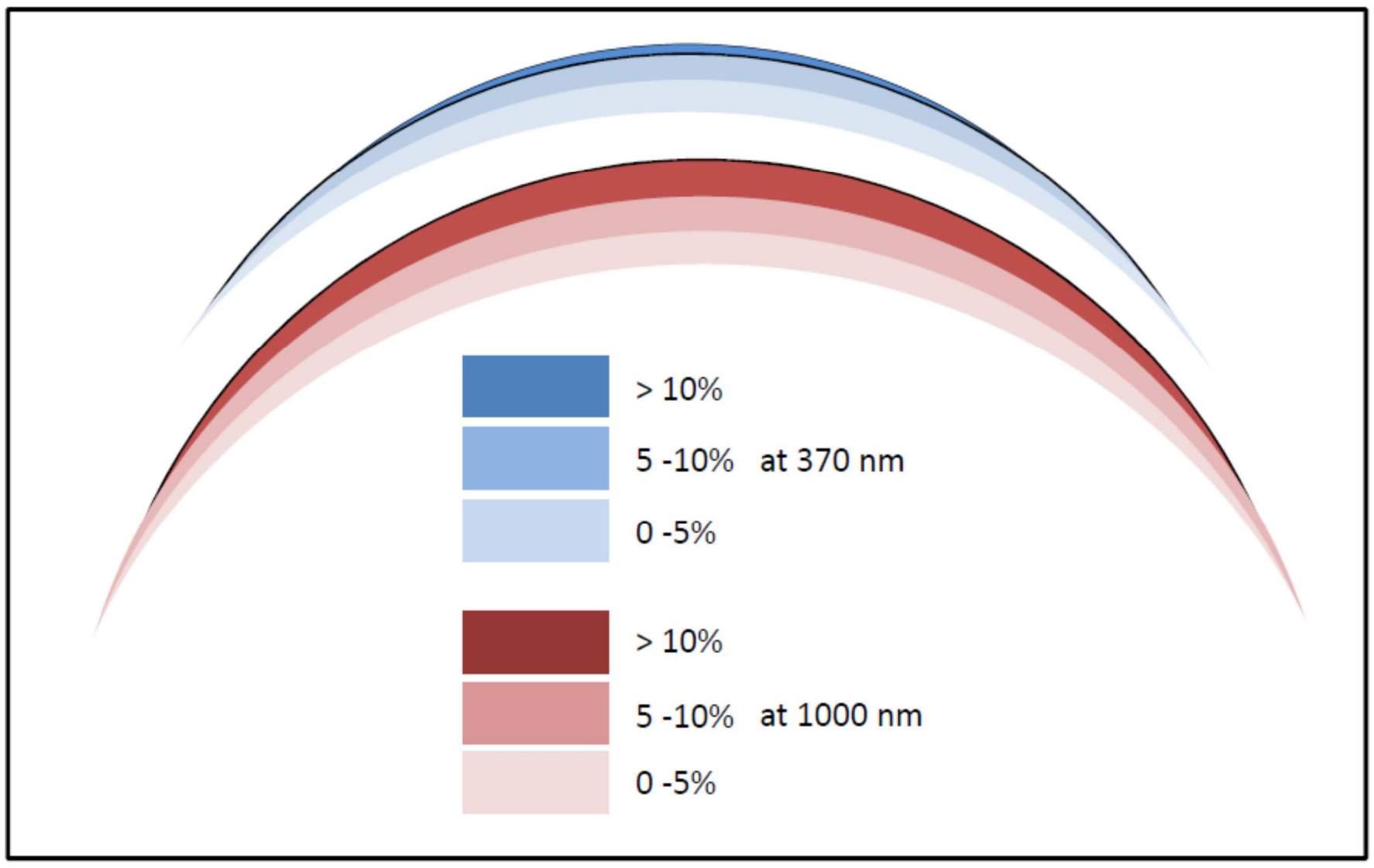

Figure 6. Vignetting at ZD=60° caused by the beam offset with respect to L1. The vignetting affects 6% of the field at 370nm, and 10% at 1000nm.

## 5. DISTORTION AND DIFFERENTIAL REFRACTION

When any wide-field telescope tracks a field, changing differential refraction causes the images to move on the focal surface. The primary distortion effect is a shear pattern compression towards the zenith, as shown in Figure 7(a). The ADC action also causes changes to the distortion pattern at the focal surface, of comparable size, but with a more complex pattern. However, the ADC action also readily allows small changes in the plate scale, through axial motion of L3 along with refocus via M2. This allows the combined distortion effects of the ADC to be tuned to partially compensate for the image motion caused by differential refraction. This works remarkably well, allowing a 3-fold reduction in the maximum image motions, and a 2-fold reduction in the mean, as shown in Figure 7(b). The pattern for other ZDs is almost identical, but scaled by tan(ZD). The benefit comes for free, just requiring very small alterations to the profiles of the surfaces used to guide the lens motions.

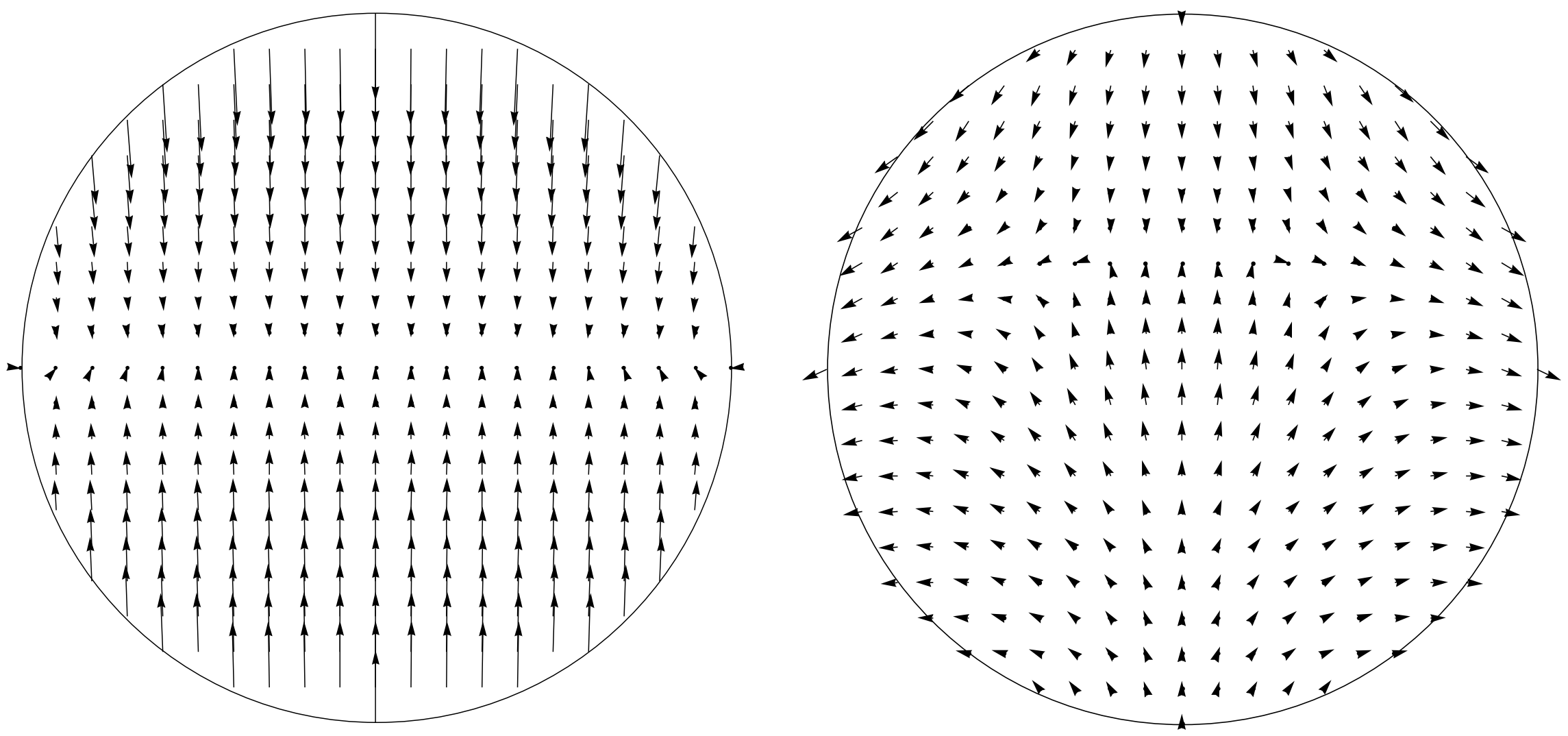

**Figure 7. Overall change in the distortion pattern for 20′ FoV between ZD=0 and ZD=45°, (a) just as caused by differential refraction, with longest vector 78μm and average 30μm, and (b) the combined effect of differential refraction and ADC action, with longest vector 26μm and average 15μm. Vectors are exaggerated in length by a factor 1000. Field diameter is 1247mm.**

Image motion during tracking comes not just from the changing ZD, but also from the change in the parallactic angle. Figure 8 shows the average image motion in μm/hr across the 20′ FoV, (a) just for differential refraction, and (b) when combined with the ADC action. The gain is a factor almost 2 at ZD<45°, with residual motions < 10μm/hr.

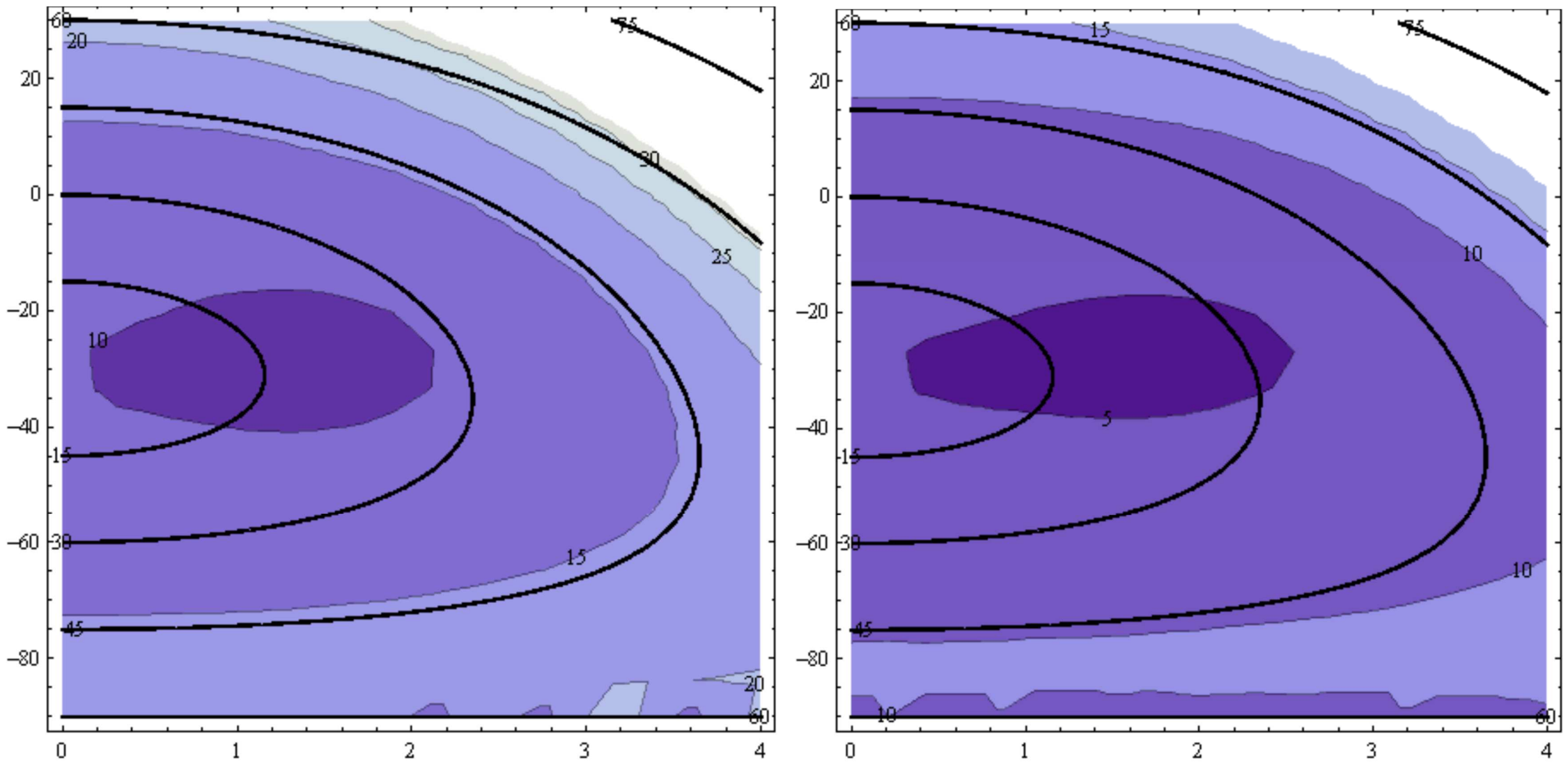

**Figure 8. The average rate of change of distortion (μm/hr) over the 20′ field, (a) caused by the ADC action, and (b) by differential refraction, both as a function of Hour Angle and Declination. Also shown are contours of constant Zenith Distance (15°, 30°, 45°, 60°).**

## 6. TELECENTRICITY

For a 20′ field at Zenith, the maximum telecentricity error (the chief ray deviation from focal surface normal) at 500nm is 0.04°. There is a large chromatic effect caused by the strongly-powered singlet L3, giving a maximum chief ray error of 0.37° for 370nm-1000nm. However, this has almost no effect on the required instrument acceptance speed, because of M1/M2 vignetting, and because the beam becomes slower off-axis. Figure 9(a) shows the illumination of the exit pupil, with the circle showing the diameter which just encloses all rays from an on-axis target at Zenith, corresponding to a speed of F/8.28. This speed suffices at all field positions and colors.

Away from Zenith, the ADC action introduces additional telecentricity errors, varying like tan(ZD), and amounting to 0.80° at ZD=60°. Figure 9(b) shows the exit pupil illumination for ZD 60°, showing that the additional non-telecentricity causes negligible additional light-loss. The design is then effectively pupil-centric.

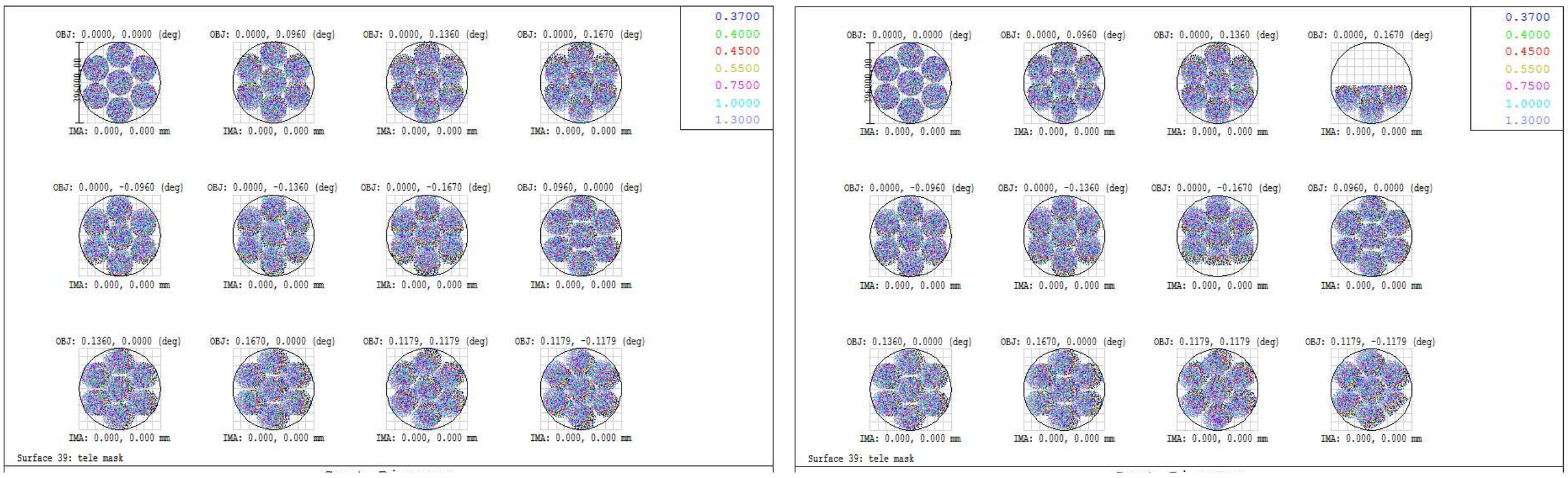

**Figure 9. Exit pupil illumination for 20′ FoV at (a) ZD=0 and (b) 50°, for all wavelengths and fields. The circle represents a speed of f/8.28, the minimum speed accepting all light from an on-axis field at Zenith. By far the largest contribution to the pupil shift comes from chromatism in L3, which means the design can only be pupil-centric at a single color at large field radii. Figure 9(b) also shows the vignetting caused by L1 and L3.**

For 10′ FoV, the telecentricity requirements are more stringent, because pupil-relaying (e.g. cold-stopping) may be required. Figure 10 shows the pupil-illumination for 10′ FoV, with variations still dominated by the chromatic effect from L3, and showing even smaller pupil shifts across the field, or with varying ZD.

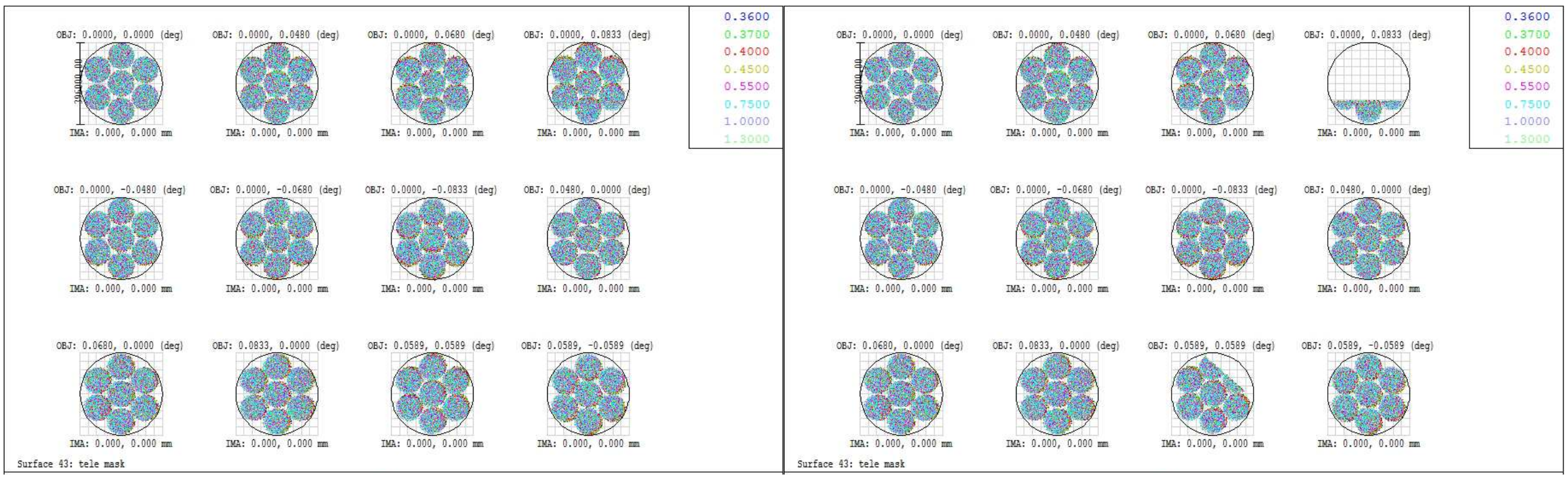

**Figure 10. Exit pupil illumination for 10′ FoV at (a) ZD=0 and (b) 50°, for all wavelengths and fields. The circle represents a speed of f/8.28, the minimum speed accepting all light from an on-axis field at Zenith. The largest non-telecentric contribution is again the chromatic effect of L3, and again the design is effectively pupil-centric.**

## 7. GUIDING AND ACTIVE OPTICS CORRECTION

The moving lens ADC design raises various issues for the Active Optical (AcO) system. Those identified to date are (a) whether the Active Guiding and Wavefront Sensing (AGWS) system can function properly, in each of its three modes, in the presence of a ZD-dependent wavefront tilt, (b) what changes will be needed in the way the AGWS system is used, and (c) whether the AcO system will then respond in the appropriate way to the AWGS information. A significant complication is that the AGWS system has pickoff mirrors above L3. Thus, although the C-ADC design is almost telecentric, the AGWS system will see very large and variable non-telecentricities, which must be accounted for before AcO correction.

To function correctly, the AGWS units should see the entire pupil (M1). The progressive vignetting caused by the ADC action restricts the useful FoV of the guiders, as seen in Figure 6. However, the guider FoV is very large, and some vignetting of the pupil is already inevitable, because M2 is only sized for on-axis use, and the AGWS units are at large field radii.

The moving lens ADC design requires a ZD-dependent correction to the telescope pointing, of up to 11″ at ZD=60°, to be included in the pointing model. Suppose an AGWS unit is then sent to an offset position corresponding to a guide star, using the standard distortion model for the telescope. In segment guiding and wavefront sensing modes, the global non-telecentricity causes the image of M1 at the AGWS pupil plane to be in error. The maximum global non-telecentricity is 0.82° at ZD=60°, causing an error of 0.7mm in the pupil position (Figure 11(a)). This can be corrected by an adjustment to the tilts of the 2 AGWS steering mirrors, to tilt the beam down by the same angle, and center the pupil image correctly (Figure 11(b)). An additional lateral offset of 2.5mm puts the image of the guide star on the center of the detector, and the S-H spots in their normal positions in both Segment Guiding and Wavefront Sensing modes (Figure 12). If the same offsets in angle and position are used for the entire field, and assuming naïve models for the radial non-telecentricity of the system, then the pupil images are correct to within 30µm, comparable with the overall error budget. A slightly more refined optical model would render the error negligible.

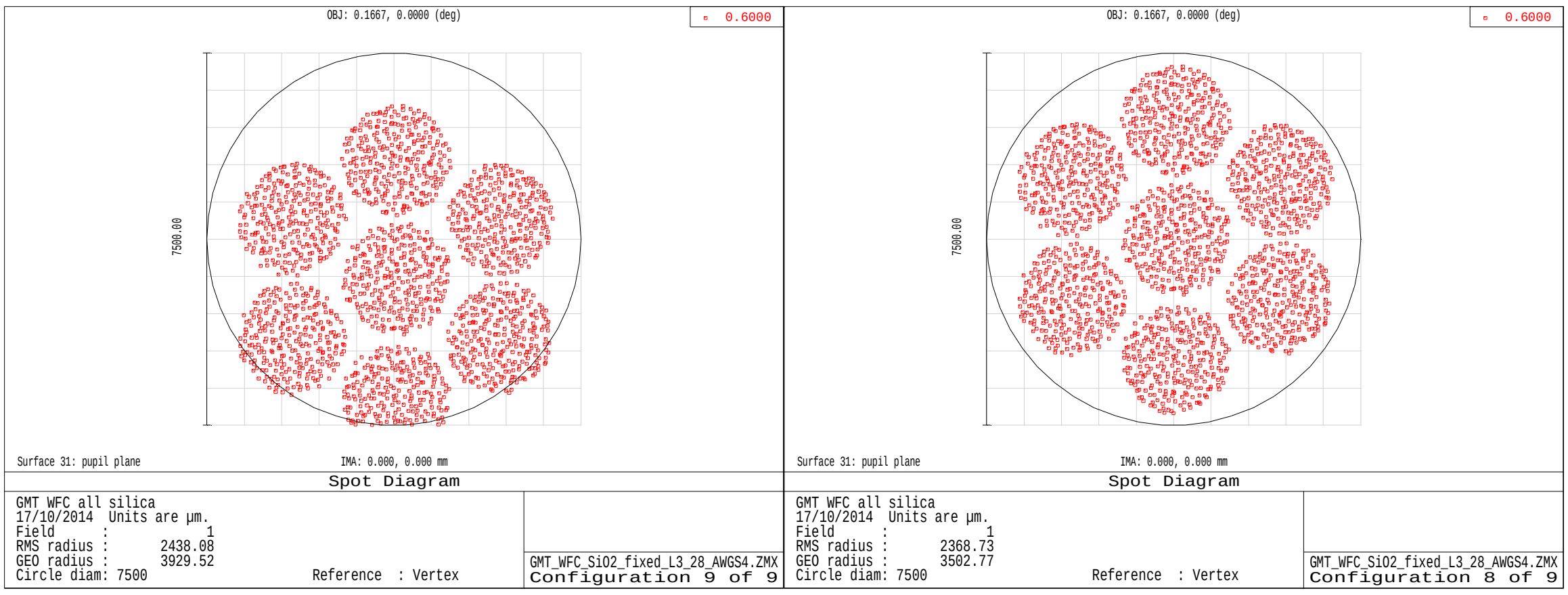

**Figure 11. Pupil image at the AGWS pupil-plane when at ZD=60°, (a) without and (b) with the AGWS offsets described in the text.**

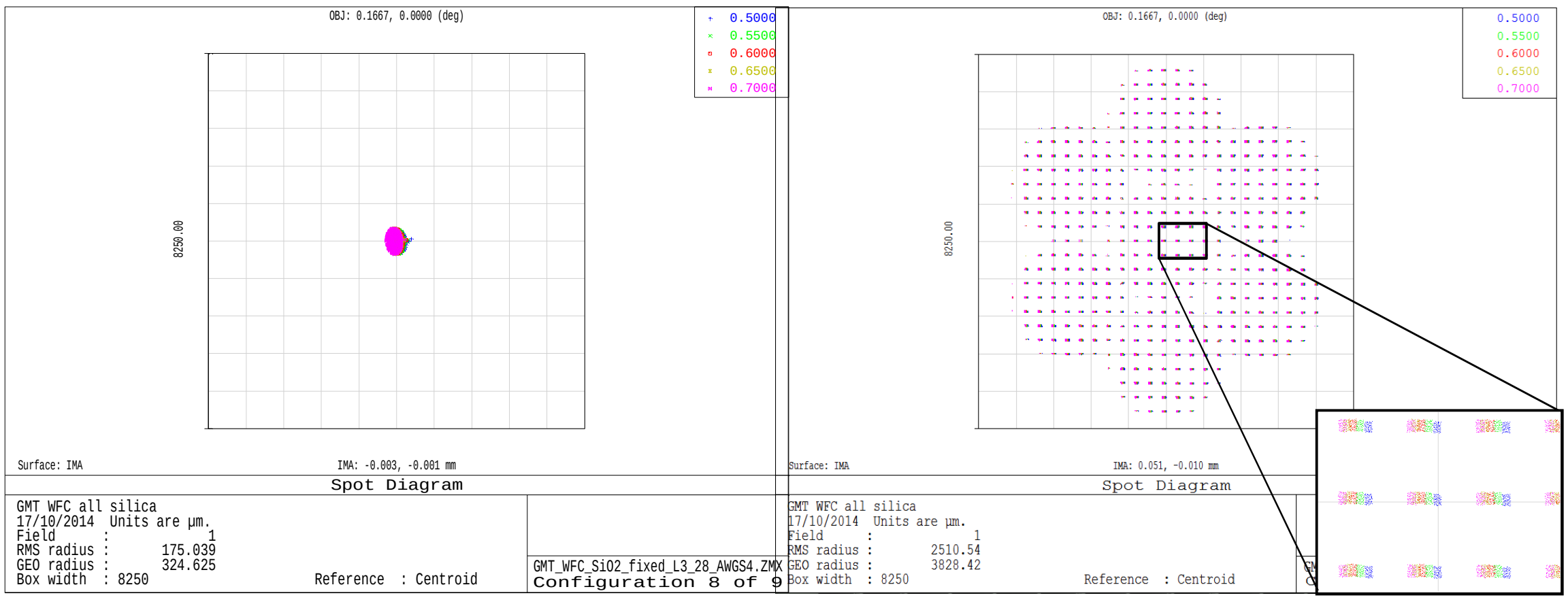

**Figure 12. (a) Guider image and (b) Wavefront Sensing Shack-Hartmann image when at ZD=60°, after AGWS offsets have been applied; the inset shows a close-up of the central Shack-Hartmann spots, showing the chromatic errors caused by the lack of L3 in the AWGS optics train.**

A complication with this approach is that as the telescope tracks, the Gregorian Instrument Rotator rotates with respect to the elevation direction. This means that the demanded offsets and mirror tilts for the AWGS system slowly change with time. The required rates of change are shown in Figure 13, and for all sensible observing they are less than 1μm/min and 1″/min. These rates of change of the position offsets are comparable to those which will be required in any case because of differential refraction.

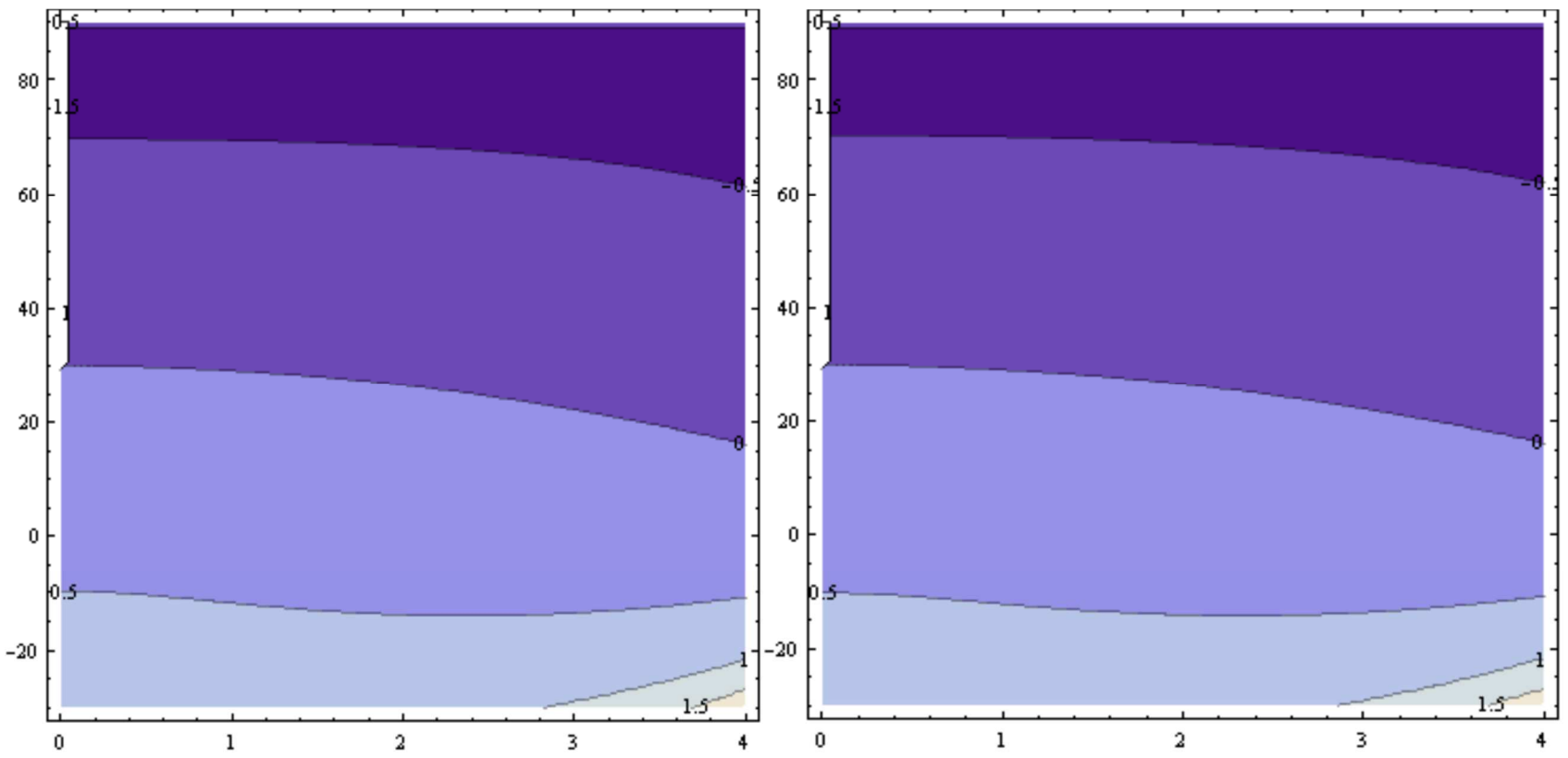

**Figure 13. Required rate of change of AGWS offset in (a) position (μm/min) and (b) angle (″/min) as a function of position ($HA, \delta$).**

The second issue is how the AcO system will respond to the changes in global focus, coma and astigmatism caused by the actuator-controlled movements of L1 and L2. The AcO system will try and correct for changes by moving M2 axially, laterally, and in tilt. These movements are essential to the image quality. As long as all the demanded movements are well within the limits of the AcO system, and insensitive to the difference in waveband between the AGWS and science observations, then there is no need for any changes to the AcO operation at this macroscopic level. Only a crude test of this has been performed. In ZEMAX, the wavelengths were restricted to a nominal guiding waveband (in the range 500-800nm), and re-optimisation was performed with three peripheral fields, and M2 allowed to move freely in axial position, translation and tilt, and all other parameters fixed. The preferred M2 movements were always very close (within ~50μm) to those determined from the full optimization. So the demanded M2 movements appear to be stable.

## 8. COATINGS AND THROUGHPUT

The design has a total thickness of 365mm of silica, giving negligible absorption throughout the optical range. Coatings will be challenging for such large lenses, with wide wavelength range and in the case of L3, strong curvature. The obvious solution is spun-coated solgel over $MgF_2$, with binding and stripping layers as needed [9], though this has only been undertaken for optics up to 1m diameter until now. It gives excellent throughput, tuneable for two wavelengths, with a smooth roll-off to short and long wavelengths. Figure 14 shows the theoretical throughput, assuming a simple solgel + $MgF_2$ coating, tuned for 400nm+800nm. The theoretical through of the C-ADC is >90% for 370nm-1800nm.

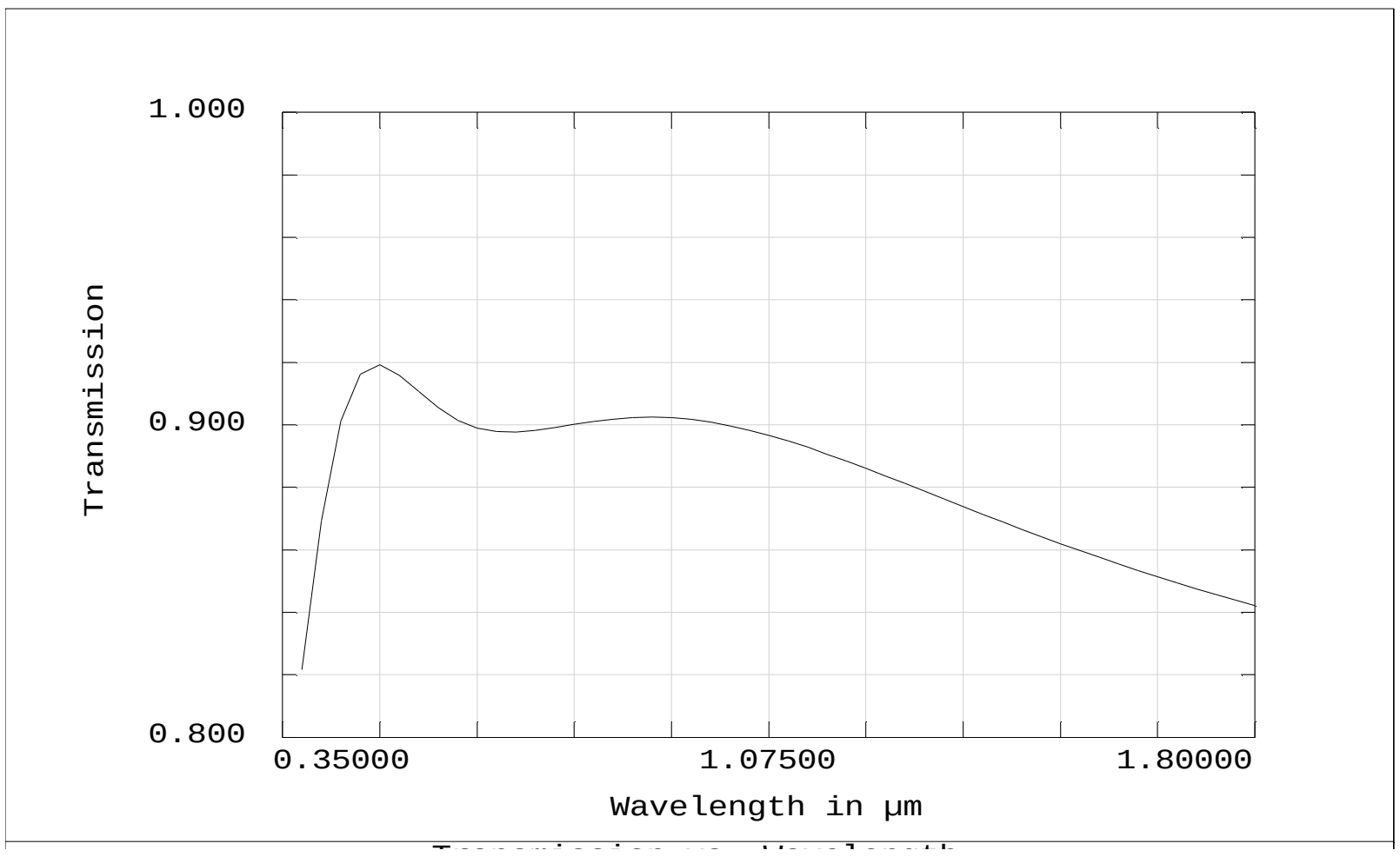

**Figure 14. Theoretical throughput for the C-ADC design, assuming solgel+$MgF_2$ coatings, and not including vignetting.**

## 9. LENS CONSIDERATIONS

The finished glass diameters and weights for the proposed designs are shown in Table 1. The diameters include an additional 30mm for mounting. The largest lens is 1510mm diameter, smaller than the largest lens proposed for LSST (1550mm). The edge thicknesses are 31mm for L1 and 29mm for L3, similar to LSST. A finite element analysis has demonstrated that these are adequate for both structural strength and optical performance in the presence of gravitational sag, but thicknesses can be increased if the polishing or handling requires it, with little penalty in performance. The overall glass mass is halved from the baseline design.

**Table 1. Lens sizes and weights**

| Lens | Diameter mm | Material | Mass (kg) |
|---|---|---|---|
| L1 | 1510 | Fused silica | 285 |
| L2 | 1440 | Fused silica | 351 |
| L3 20′ FoV | 1340 | Fused silica | 371 |
| L3 10′ FoV | 686 | Fused silica | 46 |
| **Total 20′** | | | **1007** |
| **Total 10′** | | | **681** |

Blanks are available for all lenses in the required homogeneity (±2ppm for L1 and L2, much looser for L3). Both L1 and L2 are plain spherical lenses, so polishing should not be particularly difficult or risky. L3 is aspheric in both the baseline and this design. The aspheric surface is convex, and the required asphericity is quite strong at 22 mrad maximum aspheric slope with 2.5mm removed material. But, as discussed in Section 11, the required precision on the surface is very low, since it is so near to focus in such a slow beam.

## 10. PROPOSED ADC MECHANISM

For L1 and L2, there are various ways to implement the combined lateral offset and tilts. The concept presented here which has the virtues of simplicity, accuracy, and ruggedness, requiring only one mechanism for each lens. The concept shown in Figure 15 is to have a set of 5 cam-followers, mounted from the lens barrel, running in slots mounted on the lens cell. The lateral motion is provided by an encoded actuator mounted between the barrel and lens cell. The actuator and cam-followers provide the support for the lens, while the slot profiles constrain the lens motion to the required trajectory, without over-constraint. The maximum force on the actuator is ~4000N.

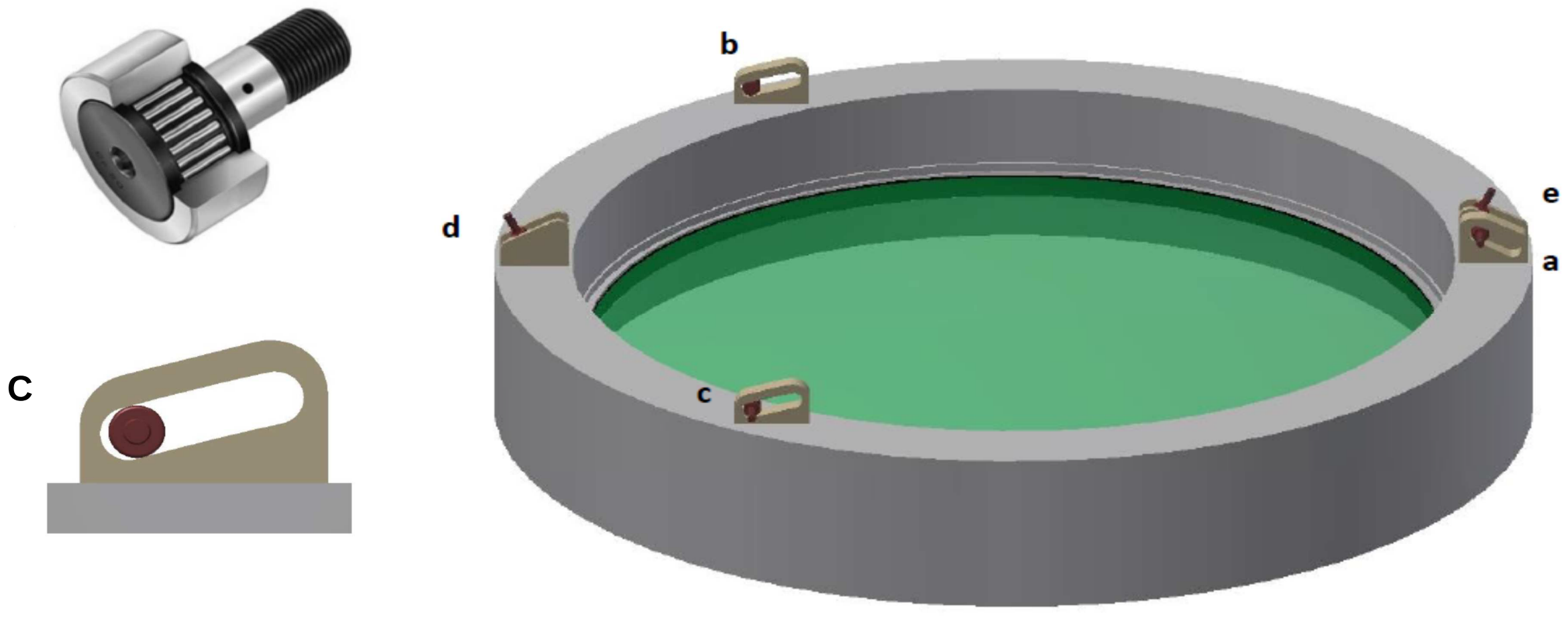

**Figure 15. Proposed support system for L1, with 5 cam-followers (as shown upper left) mounted from the WFC barrel, running in slots (e.g. as shown lower left) on the lens cell and giving the required trajectory. A single actuator (not shown) would push the lens cell from right to left to give the ADC action.**

This arrangement won't work for L3, which is mounted below the instrument derotator. Thus the shift and tilt has to be in a direction that changes continuously with respect to the lens. Also, the lens is mounted in a confined space. However, we have found a simple design that gives the required support and motion, without over-constraint, as shown in Figure 15. The lens cell has a toroidal surface, and rests on three tracker balls. Three actuators are also mounted between the instrument structure and the top surface of the lens cell, but not radially. Then rotation of the lens by small amounts is possible, avoiding any over-constraint. The toroidal surface is to allow the desired axial motion of L3 as it translates and tilts.

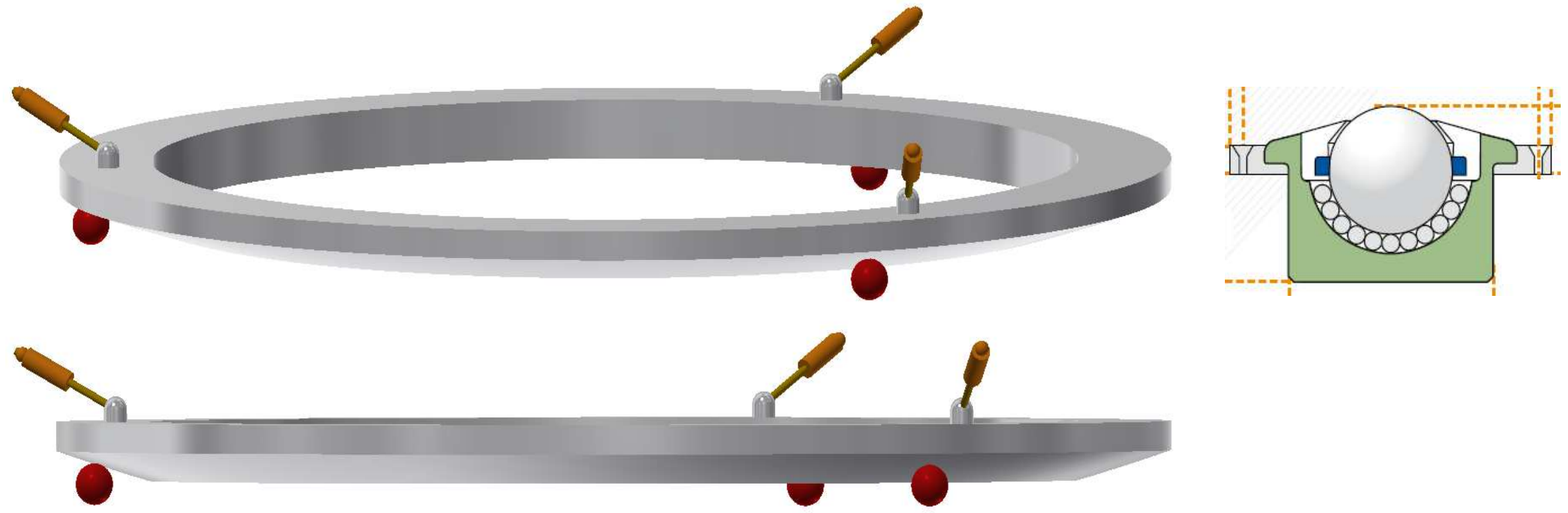

Figure14. Two views showing essentials of the offset/tilt mechanism for L3, and a section through a tracker ball.

## 11. TOLERANCES

A full tolerance analysis has not yet been carried out. However, some aspects that seemed critical have been investigated. Overall, we found that (a) the system is very tolerant to axial shifts of L1 and L2; (b) the tilt/translation ratio for L1 and L2 is critical, with variations of more than ~1% causing significant degradation, and (c) given the correct tilt/translation ratio, the required absolute precision of the tilts and translations is very low, with 1% errors having negligible effect.

With L1 and L2 being of fused silica and without aspheres, their surface quality and homogeneity requirements are unlikely to be very testing, at least compared with other large WFCs such as HSC or LSST. The small changes in the position of the beam with respect to the lenses (maximum of a few 10s of mm/hour) avoid the very extreme homogeneity requirements of Amici ADCs.

L3 is a reasonably strong asphere, with slope deviation up to 22mrad and 2.5mm of removed material. However, it is very close to focus, making it very tolerant to surface and homogeneity errors. A sensitivity test in ZEMAX showed that even 10μm peak to valley mid-range surface errors would be tolerable, and more realistic few-wave errors would have negligible effect.

## 12. DESIGN VARIANTS

### L1/L2 movement options

The ADC action depends on the relative translations of L1 and L2, but there is almost complete freedom in their absolute values – e.g., all the translation could be on L2. For any pair of translations, aberration control then calls for precise (and sadly always non-zero) tilts of both lenses. We have chosen to split the translation equally between the two lenses, since this minimises the sizes of the proposed actuator mechanism. This choice also has the feature that the beam moves hardly at all with respect to L2, so there are no vignetting losses there.

### L3/focal surface movement

Earlier versions of the design had L3 fixed. This causes a small degradation of image quality, and introduced a significant non-telecentricity at large ZDs, causing problems with the pupil-relaying and baffling at M2. An intriguing but rejected option is that tilting the instrument along with L3 gives superb monochromatic image quality at all ZDs.

### Glass Choices

Early versions of the design had LLF1 for L1 and L2. This gave somewhat worse imaging at the zenith ($d_{80} < 0.060''$), but much better ADC performance ($d_{80} < 0.127''$ at ZD=60°), as the partial dispersion of LLF1 is a better match than fused silica to the atmosphere. There is a throughput hit of a few %, rising to 15% at 370nm. LLF6 is also suitable but no longer listed as a preferred glass for large lenses [10].

L3 also has a glass choice, it could be of fused silica, N-BK7 or N-FK5. N-BK7 gives the best imaging, but the worst transmission; it requires a significantly less steep asphere than silica. The closeness to focus makes it very tolerant to glass inhomogeneity. N-FK5 would be an excellent material for L3, since it would reduce the chromatic non-telecentricity, but we have been advised that it is unlikely to be made available in this size, despite the claims in [10].